\documentclass
[letterpaper,english,pra,aps,preprint,nofootinbib,superscriptaddress]{revtex4}%
\usepackage{graphicx}
\usepackage{color}
\usepackage{bm}
\usepackage{amsmath}
\usepackage{amssymb}
\usepackage{amsfonts}%

\begin{document}
\title{Null weak values and the past of a quantum particle}
\author{Q.\ Duprey}
\affiliation{Laboratoire de Physique Th\'{e}orique et Mod\'{e}lisation (CNRS Unit\'{e}
8089), Universit\'{e} de Cergy-Pontoise, 95302 Cergy-Pontoise cedex, France}
\author{A. Matzkin}
\affiliation{Laboratoire de Physique Th\'{e}orique et Mod\'{e}lisation (CNRS Unit\'{e}
8089), Universit\'{e} de Cergy-Pontoise, 95302 Cergy-Pontoise cedex, France}
\affiliation{Institute for Quantum Studies, Chapman University, Orange, CA 92866, United
States of America}

\begin{abstract}
Non-destructive weak measurements (WM) made on a quantum particle allow to
extract information as the particle evolves from a prepared state to a finally
detected state. The physical meaning of this information has been open to
debate, particularly in view of the apparent discontinuous trajectories of the
particle recorded by WM. In this work we investigate the properties of
vanishing weak values for projection operators as well as general observables.
We then analyze the implications when inferring the past of a quantum
particle. We provide a novel (non-optical) example for which apparent
discontinuous trajectories are obtained by WM.\ Our approach is compared to
previous results.

\end{abstract}
\maketitle

\section{\label{intro}Introduction}

Assume a quantum system is prepared in some initial state at time
$t_{i}$, and ultimately detected and found to be in some final state
at time $t_{f}$. It is usually taken for granted that quantum
mechanics does not allow to learn anything concerning the property
of the system at some intermediate time. The reason is that in order
to learn something about a given property, the associated observable
needs to be measured. But as is well-known, measurements are special
in quantum mechanics: measurements break the unitary evolution and
project the premeasurement system state to one of the eigenstates of
the measured observable. Hence in typical cases a measurement made
at some intermediate time will irremediably disturb the system
evolution from what it would have been without this intermediate
measurement. The upshot is that it is impossible to ascertain the
particle's properties, and in particular its past when the system
has evolved from a given initial state to a final state. The best we
can do is employ counterfactual reasoning, but Bohr has long ago
warned us \cite{bohr-einsteinbook} that this would lead to
paradoxes, as exemplified in the well-known Delayed Choice
Experiment proposed by Wheeler \cite{wheeler-delayed}.

However there have been recent proposals to ascertain the paths taken by a
quantum particle. In particular Vaidman examined the path of a photon in
nested interferometers \cite{vaidman2013}, while one of us investigated the
dynamical paths compatible with a given final state when a quantum system
evolution is generated by a semiclassical Feynman propagator \cite{A2012PRL}.
These proposals are based on weak measurements. Weak measurements were
introduced \cite{AAV} in 1988 as a theoretical scheme for minimally perturbing
non-destructive quantum measurements. Aharonov, Albert and Vaidman precisely
showed \cite{AAV} that, without departing from the standard quantum formalism,
it was possible to measure an observable $A$ in a particular sense without
appreciably changing the system evolution. The main idea is to achieve an
interaction with a weak coupling between $A$ and a dynamical variable of an
external degree of freedom (an ancilla that will be called ``quantum
pointer''). The system and the quantum pointer are entangled, until the final
projective measurement of a different system observable $B$ correlates the
obtained system eigenstate with the quantum state of the weak pointer. The
state of the weak pointer has picked up a shift (relative to its initial
state) proportional to a quantity known as the \textit{weak value } of $A$.
When a weak value vanishes, the state of the quantum pointer remains
unchanged, and Refs \cite{vaidman2013,A2012PRL} interpreted this fact by
asserting that the system property coupled to the pointer was not there
(otherwise the pointer state would have changed).

While many experimental and theoretical works dealing with weak measurements
have been published in the last decade (see \cite{RMP} for a review) the
meaning of the observed weak values has been debated since their inception,
from the early comments by Leggett \cite{leggett89} and Peres \cite{peresWM}
to more recent works \cite{sokolovski16,svensson2014}. Unsurprisingly, any
proposal to infer the past of a quantum system from the weak values is going
to face criticism disputing the relevance of weak measurements concerning the
properties that can be ascribed to a system during its evolution. In
particular, Vaidman \cite{vaidman2013,A2012PRL} noted that the weak values of
the spatial projector were non-zero inside a Mach-Zehnder interferometer (MZI)
inserted on one of the arms of another larger Mach-Zehnder, but the weak
values along that arm did vanish before and beyond the nested MZI. The same
feature was also remarked \cite{A2013JPA} in a 3 path interferometer: when 2
of the 3 branches are joined, the spatial projector weak value (that did not
vanish on either of these 2 arms) vanishes once these 2 branches merge. If a
non-vanishing weak value is interpreted as a trace left by a particle, while a
vanishing weak value implies the particle wasn't there, one would be led to
conclude for instance that the particle was inside the nested MZI while it
could never have entered or exited, a rather strange conclusion.

Indeed, several authors
\cite{zubairy2013,vaidman-com2013,saldanha2014,bartkiewicz2015,ferenczi2015,jordan-dove,sokolovskiMZI2016,griffiths2016,bula2016,vaidman-com2016}
have criticised such an idea, generally basing their critcism on the
experimental realization \cite{danan2013} of Vaidman's nested MZI proposal.
Some of the criticism \cite{zubairy2013,saldanha2014,ferenczi2015,bula2016} is
essentially relevant to the details of the experiment (that employed tilting
mirrors and classical electromagnetic waves). In this paper we will instead be
concerned by fundamental issues concerning the properties of a quantum system
between preparation and detection. Indeed, relying on a classical optics
experiment, or even on quantum optics, in order to interpret a quantity
derived in the context of non-relativistic quantum mechanics requires at best
an amount of extrapolation that will not help in giving a solid account of the
meaning of null weak values. This is precisely the aim of the present work: to
analyze and understand null weak values, and from there examine which
interpretations can make sense. To the best of our knowledge, such a work has
not been undertaken.

This work is organized as follows. We will first recall the weak measurements
formalism (Sec. II). We will then carefully scrutinize the case of vanishing
weak values and give a couple of illustrations (Sec. III). Sec. IV will be
devoted to the interpretation of null weak values, and we will discuss and
compare with the views expounded in the recent papers
\cite{vaidman2013,saldanha2014,vaidman-past,jordan-dove,sokolovskiMZI2016,griffiths2016}%
. We will draw our conclusions in Sec. V.

\section{Weak Measurements\label{wv-f}}

The underlying idea at the basis of the weak measurement (WM) framework is to
give an answer the question:\textquotedblleft\textit{what is the value of a
property (represented by an observable $A$) of a quantum system while it is
evolving from an initial state }$\mathit{\left\vert \psi(t_{i})\right\rangle
}$\textit{\ to a final state $\left\vert \chi(t_{f})\right\rangle $ obtained
as a result of an usual projective measurement?}\textquotedblright. As our
interest in this paper concerns the instance of null weak values, we will
restrict our exposition to the simplest case, a bivalued observable $A$, with
eigenstates and eigenvalues denoted by $A\left\vert a_{k}\right\rangle
=a_{k}\left\vert a_{k}\right\rangle ,$ $k=1,2$.

Let us assume that at $t=t_{i}$ the system of interest is prepared into the
state $\left\vert \psi(t_{i})\right\rangle $ (this step is known as
preselection).\ An ancilla (that will play the role of a quantum pointer) is
at that time in state $\left\vert \varphi(t_{i})\right\rangle ,$ so the total
initial quantum state is the uncoupled state
\begin{equation}
\left\vert \Psi(t_{i})\right\rangle =\left\vert \psi(t_{i})\right\rangle
\left\vert \varphi(t_{i})\right\rangle . \label{inis}%
\end{equation}
We assume the system and the pointer will interact during a brief time
interval $\tau$ centered around $t=t_{w}$ (physically corresponding to the
time during which the system and the quantum pointer interact). Let the
interaction Hamiltonian be specified by
\begin{equation}
H_{int}=g(t)AP \label{Hint}%
\end{equation}
coupling the system observable ${A}$ to the momentum $P$ of the
pointer.
$g(t)$ is a smooth function non-vanishing only in the interval $t_{w}%
-\tau/2<t<t_{w}+\tau/2$ and such that $g\equiv\int_{t_{w}-\tau/2}^{t_{w}%
+\tau/2}g(t)dt$ appears as the effective coupling constant. Eq. (\ref{Hint})
is nothing but the usual interaction employed to account for projective
measurements of $A$ : in that case $g(t)$ is a sharply peaked function
correlating each $\left\vert a_{k}\right\rangle $ to an orthogonal state of a
macroscopic pointer, that collapses projecting the system state to a random
eigenstate $\left\vert a_{k_{0}}\right\rangle $. Here instead $g(t)$ will be
small, the pointer is quantum, and the pointer-system will evolve unitarily
until a subsequent projective measurement made on the system will correlate
the quantum pointer to a specific final state of the system, as we now detail.

Let us denote by $U(t_{w},t_{i})$ the system evolution operator between
$t_{i}$ and $t_{w}$ and disregard the self-evolution of the pointer state.
After the interaction $(t>t_{w}+\tau/2)$ the initial uncoupled state
(\ref{inis}) has become entangled:
\begin{align}
\left\vert \Psi(t)\right\rangle  &  =U(t,t_{w})e^{-ig{A}P}U(t_{w}%
,t_{i})\left\vert \psi(t_{i})\right\rangle \left\vert \varphi(t_{i}%
)\right\rangle \label{corrs1}\\
&  =U(t,t_{w})e^{-ig{A}P}\left\vert \psi(t_{w})\right\rangle
\left\vert
\varphi(t_{i})\right\rangle \label{corrsl2}\\
&  =U(t,t_{w})\sum_{k=1,2}e^{-iga_{k}P}\left\langle a_{k}\right\vert \left.
\psi(t_{w})\right\rangle \left\vert a_{k}\right\rangle \left\vert
\varphi(t_{i})\right\rangle . \label{corrs}%
\end{align}
Finally, the system undergoes a standard projective measurement at time
$t_{f}:$ an observable $B$ (different from $A$) is measured and the system
ends up in one of its eigenstates $\left\vert b_{k}\right\rangle $. Let us
only keep the results corresponding to a chosen eigenvalue $b_{k_{0}}$ and
label the postselected state by $\left\vert \chi(t_{f})\right\rangle
\equiv\left\vert b_{k_{0}}\right\rangle $.\ The projection on the entangled
state $\left\vert \Psi(t_{f})\right\rangle $ given by Eq. (\ref{corrs}) leads
to the final state of the pointer correlated with the postselected system
state:
\begin{equation}
\left\vert \varphi(t_{f})\right\rangle =\sum_{k=1,2}\left[  \left\langle
\chi(t_{w})\right\vert \left.  a_{k}\right\rangle \left\langle a_{k}%
\right\vert \left.  \psi(t_{w})\right\rangle \right]  e^{-iga_{k}P}\left\vert
\varphi(t_{i})\right\rangle . \label{finalps}%
\end{equation}
If $\left\vert \varphi(t_{i})\right\rangle $ is a localized state in the
position representation, then $\varphi(x,t_{f})$ is given by a superposition
of shifted initial states
\begin{equation}
\varphi(x,t_{f})=\sum_{k=1,2}\left[  \left\langle \chi(t_{w})\right\vert
\left.  a_{k}\right\rangle \left\langle a_{k}\right\vert \left.  \psi
(t_{w})\right\rangle \right]  \varphi(x+ga_{k},t_{i}). \label{finalpspos}%
\end{equation}
This expression is the first step of the usual von Neumann projective by which
each eigenstate $\left\vert a_{k}\right\rangle $ of the measured observable is
correlated with a given state $\varphi(x+ga_{k})$ of the pointer (but in a von
Neumann measurement the second step is a projection to an eigenstate
$\left\vert a_{k_{f}}\right\rangle $ of A, which does not happen here).

Let us now assume the coupling $g$ is sufficiently small so that
$e^{-iga_{k}P}\approx1-iga_{k}P$ holds for each $k$. Eq. (\ref{finalps})
becomes
\begin{align}
\left\vert \varphi(t_{f})\right\rangle  &  =\left\langle \chi(t_{w}%
)\right\vert \left.  \psi(t_{w})\right\rangle \left(
1-igP\frac{\left\langle \chi(t_{w})\right\vert {A}\left\vert
\psi(t_{w})\right\rangle }{\left\langle \chi(t_{w})\right\vert
\left.  \psi(t_{w})\right\rangle
}\right)  \left\vert \varphi(t_{i})\right\rangle \\
&  =\left\langle \chi(t_{w})\right\vert \left.  \psi(t_{w})\right\rangle
\exp\left(  -igA^{w}P\right)  \left\vert \varphi(t_{i})\right\rangle
\label{finwv}%
\end{align}
where
\begin{equation}
A^{w}=\frac{\left\langle \chi(t_{w})\right\vert A\left\vert \psi
(t_{w})\right\rangle }{\left\langle \chi(t_{w})\right\vert \left.  \psi
(t_{w})\right\rangle } \label{wvt}%
\end{equation}
is the weak value of the observable $A$ given pre and postselected states
$\left\vert \psi\right\rangle $ and $\left\vert \chi\right\rangle $
respectively (we will sometimes employ instead the full notation
$A_{\left\langle \chi\right\vert ,\left\vert \psi\right\rangle }^{w}$ to
specify pre and postselection). For a localized pointer state, expanding to
first order the terms $\varphi(x+ga_{k},t_{i})$ in Eq. (\ref{finalpspos})
leads to Eq. (\ref{finwv}): the overall shift $\varphi(x+gA_{w},t_{i})$ is
readily seen to result from the interference due to the superposition of the
slightly shifted terms $\varphi(x+ga_{k},t_{i})$.

We can now summarize the weak measurement protocol: (i) preselection, ie
preparation of the initial state (\ref{inis}); (ii) weak coupling through the
measurement Hamiltonian (\ref{Hint}); (iii) postselection, leading to the
quantum state of the pointer (\ref{finwv}); (iv) readout (measurement) of the
quantum pointer. The quantum pointer readout allows to extract the weak value:
Eq. (\ref{finwv}) indicates that the pointer will undergo a translation
proportional to the weak value.

\section{Null Weak Values}

\subsection{Weak values: general properties}

Following Eqs. (\ref{finalpspos})-(\ref{finwv}) the real part of the weak
value $A_{w}$ appears as the shift brought to the average initial pointer
state position $\varphi(x,t_{i})$ due to its coupling with the system via the
local interaction Hamitonian (\ref{Hint}). The weak values are generally
different from the eigenvalues. Indeed, the weak coupling step correlates the
system observable eigenstates with pointer states, but a single eigenstate is
only obtained for strong couplings (relative to the pointer states spread),
and that a random projection takes place. The system's state is thus radically
modified when undergoing a transition from its pre-measurement state to an
eigenstate. The eigenvalue associated to this observable eigenstate reflects
the value taken by the corresponding property after this radical change of state.

Instead, the system-pointer coupling in a weak measurement practically leaves
the system state unaffected: since%
\begin{equation}
e^{-ig{A}P}\left\vert \psi(t_{w})\right\rangle \left\vert \varphi
(t_{i})\right\rangle \approx\left\vert \psi(t_{w})\right\rangle
\left\vert \varphi(t_{i})\right\rangle -iA\left(  g\left\vert
\psi(t_{w})\right\rangle \right) P \left\vert
\varphi(t_{i})\right\rangle
\end{equation}
The tiny fraction $g\left\vert \psi(t_{w})\right\rangle $ of the system state
that interacts is precisely the one that couples to the quantum pointer. The
weak value appears as the imprint of this coupling left on the pointer,
conditioned on the final projective measurement (postselection\footnote{Of
course postselection irremediably modifies the system state, as per any
projective measurement.}). The weak value as defined from Eq. (\ref{wvt}) can
be seen as the ratio of the transition amplitude to the final state
$\left\vert \chi(t_{w})\right\rangle $ of the fraction of the state
$A\left\vert \psi(t_{w})\right\rangle $ that has interacted relative to a
non-interaction situation in which the system state remains $\left\vert
\psi(t_{w})\right\rangle $. In particular the numerator is the standard
transition amplitude matrix element for the observable $A.$ Hence a weak value
cannot be associated with an eigenstate but with a transition from a
preselected to a postselected state. Nevertheless, weak values obey a similar
relation with regard to the computation of expectation values: the standard
expectation value of $A$ in state $\left\vert \psi(t_{w})\right\rangle $ is
given in terms of eigenvalues by the textbook expression
\begin{equation}
\left\langle \psi(t_{w})\right\vert {A}\left\vert
\psi(t_{w})\right\rangle
=\sum_{f}\left\vert \left\langle a_{f}\right\vert \left.  \psi(t_{w}%
)\right\rangle \right\vert ^{2}a_{f}. \label{sav}%
\end{equation}
It can also be written in terms of weak values as
\begin{equation}
\left\langle \psi(t_{w})\right\vert A\left\vert \psi(t_{w})\right\rangle
=\sum_{f}\left\vert \left\langle \chi_{f}(t_{f})\right\vert \left.  \psi
(t_{f})\right\rangle \right\vert ^{2}A_{\left\langle \chi_{f}\right\vert
,\left\vert \psi\right\rangle }^{w} \label{expvwv}%
\end{equation}
with $\left\vert \chi_{f}(t_{f})\right\rangle =\left\vert b_{f}\right\rangle
.$ Rather than involving the probability of obtaining an eigenstate, Eq.
(\ref{expvwv}) is expressed in terms of the probabilities of obtaining a
postselected state $\left\vert b_{f}\right\rangle .$ Then the weak value
indicated by the quantum pointer that was coupled to $A$ replaces the
eigenvalue in the usual expression (\ref{sav}). Note that the imaginary part
of the right handside of Eq. (\ref{expvwv}) is zero.

\subsection{Null weak values\label{nullwv}}

\subsubsection{Vanishing eigenvalues}

Let us first examine the case of vanishing \textit{eigenvalues}. In
the standard von Neumann measurement scheme, a null eigenvalue
implies that the (macroscopic) pointer state is left untouched: the
coupling has no effect on the pointer. But apart from this
specificity, a vanishing eigenvalue appears as the result of a
standard projective measurement: the system state changes, as it is
projected to the eigenstate associated with the null eigenvalue for
the measured observable.\ For example imagine a particle entering a
Mach-Zehnder interferometer. After the beamsplitter, its quantum
state of each atom can be described by the superposition $\left\vert
I\right\rangle +\left\vert II\right\rangle $, where $\left\vert
I\right\rangle $ ($\left\vert II\right\rangle $) denotes the
wavepackets traveling along arm $I$ ($II\text{)}$. If a standard
measurement of the projector onto path $\Pi _{I}\equiv\left\vert
I\right\rangle \left\langle I\right\vert $ yields 0, then (i) the
particle is not on path $I$ and (ii) its quantum state has collapsed
to $\left\vert II\right\rangle $ (one is certain to find the
particle on that path).

As another example, consider a particle with integer spin. Then measuring the
spin projection along some direction can yield a null eigenvalue.\ The spin
state is then projected to the corresponding eigenstate (as can be verified by
making subsequent measurements) corresponding to no spin component along that
direction. Hence we can assert that when a vanishing eigenvalue is obtained,
the initial system state has been radically perturbed (as per any projective
measurement) but the pointer state has remained the same because the property
that has been measured is not there (no particle, no spin component).

\subsubsection{Transition amplitudes}

As seen above, for weak measurements the system's state is not projected after
the weak coupling.\ Hence a null weak value leaves the pointer untouched (the
coupling has no effect) just as in the case of null eigenvalues, but the
implication does not concern eigenvectors but transitions to the postselected
state. This follows from the weak values definition (\ref{wvt}): $A^{w}=0$ iff
$\left\langle \chi(t_{w})\right\vert A\left\vert \psi(t_{w})\right\rangle =0$,
so a vanishing weak value is obtained when the transition between the fraction
of the state that has interacted $A\left\vert \psi(t_{w})\right\rangle $ and
the postselected state is forbidden. As explained in Sec. \ref{wv-f}, if the
evolution of the states between the initial, interaction and postselection
times is not trivial, then the vanishing transition is between the state at
the time the weak coupling takes place (with the preselected state forwarded
in time) and the postselected state evolved backward in time, or alternatively
with the transformed state $A\left\vert \psi(t_{w})\right\rangle $ evolved up
to $t_{f}$ and the postselected state.

As is well known from elementary quantum mechanics, a forbidden transition
means that the final state cannot be reached under the action of the
observable operator on the initial state. Here the final state is the
postselected state, and the action of the operator transforming the
pre-measurement state is physically due to the weak interaction between the
system and the quantum pointer. Under this setting, weak measurements can be
seen as an experimentally feasible protocol in order to measure the vanishing
transition amplitudes.

\subsubsection{Meaning of null weak values}

A null weak value correlates successful postselection with the quantum pointer
having been left unchanged despite the interaction with the system. The
reason, as seen in the preceding paragraph, is that the transition amplitude
vanishes. If the postselected state is obtained, then the property represented
by $A$ cannot be detected by the weakly coupled quantum pointer. For example
when the weak value $\Pi_{I}^{w}$ of a spatial projector $\Pi_{I}%
\equiv\left\vert I\right\rangle \left\langle I\right\vert $ vanishes this
means that the postselected state cannot be reached from the region
$\left\vert I\right\rangle $ where the weak interaction took place. So in a
sense to be specified and refined below, it is cogent to assert that the
system could not have been in region $\left\vert I\right\rangle $ (conditioned
on successful postselection) because quantum correlations prevent the system
from reaching the final state from a particle localized in that region at the
time it coupled to the pointer. For some more general observable $A$, a null
weak value $A^{w}=0$ means that the transformation produced by the coupling on
the system is such that the postselected state cannot be reached. For this
reason we may say again that the property corresponding to $A$ is "{}not
there"{} in the region where the interaction took place, consistently with the
fact that the quantum pointer's state remains unchanged by the coupling.

\subsection{Illustrations}

\subsubsection{3-path interferometer\label{3ex}}

Let us assume spin-1 particles (e.g, atoms) are separated by a beam splitter
into 3 paths. To be specific let us take the initial state as
\begin{equation}
\left\vert \psi_{i}\right\rangle =\left\vert m_{z}=0\right\rangle \left\vert
\xi\right\rangle \label{ini}%
\end{equation}
where $\xi(\mathbf{r})\equiv\left\langle \mathbf{r}\right\vert \left.
\xi\right\rangle $ is the spatial part of the wavefunction and $\left\vert
m_{z}=0\right\rangle $ stands for the spin state $\left\vert J=1,m_{z}%
=0\right\rangle $ (spin projection quantized along the $\mathbf{\hat{z}}$ axis
with azimuthal number $m_{z}=0$). We assume $\xi(\mathbf{r})$ can be
represented by a Gaussian.

\begin{figure}[tb]
\setlength{\fboxsep}{-45pt}\setlength{\fboxrule}{0pt}\fbox{\includegraphics[scale=0.17]{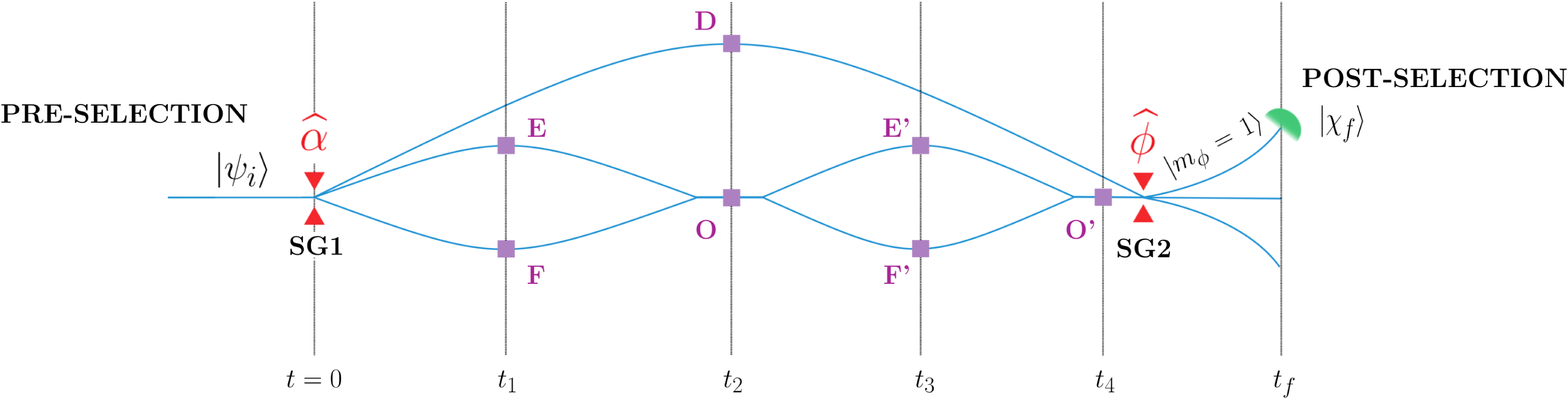}}%
\caption{A 3 path interferometer for spin-1 particles with a provision for
recombination of the 2 lower paths. Weak measurements take place at times
$t_{k}$ (as indicated at the bottom) at the points shown on the figure. For
appropriately chosen preselected and postselected states (see Text), null weak
values are obtained at $O$ and $O^{\prime}$ but not at $E$, $F$, $E^{\prime}$
or $F^{\prime}$.}%
\label{fig3path}%
\end{figure}

At $t=0$ the wavepacket enters the beamsplitter region denoted SG on Fig.
\ref{fig3path}. For $t>0$, $\left\vert \xi\right\rangle $ separates into three
wavepackets\ each associated with a given value of $m_{\alpha}=-1,0,1$, and
the wavefunction becomes
\begin{equation}
\left\vert \psi(t)\right\rangle =\sum_{k=-1,0,1}d_{k}(\alpha)\left\vert
m_{\alpha}=k\right\rangle \left\vert \xi_{k}(t)\right\rangle .\label{27}%
\end{equation}
The states $\left\vert m_{\alpha}=\pm 1,0\right\rangle $ are the
three eigenstates of the spin component along the direction
$\mathbf{\hat{\alpha}}$ and the complex numbers $d_{k}(\alpha)$ are
given by $d_{k}(\alpha )=\left\langle m_{\alpha}=k\right\vert \left.
m_{z}=0\right\rangle $ \footnote{Technically, SG is a Stern-Gerlach
apparatus with an inhomogeneous magnetic field directed along the
direction $\mathbf{\hat{\alpha}}$. This separates the wavepackets
according to their associated spin projection along
$\mathbf{\hat{\alpha}}$. $d_{k}(\alpha)$ is given by the reduced
Wigner rotational matrix element generally denoted $\left\langle
m_{\alpha
}\right\vert \left.  m_{\beta}\right\rangle \equiv d_{m_{\alpha},m_{z}}%
^{J=1}(\beta-\alpha)$ .} . The wavepackets then evolve\footnote{In
principle the dynamics of the wavepackets $\left\vert
\xi_{k}\right\rangle $ can be computed exactly by solving the
Schr\"{o}dinger equation of a particle in an inhomogeneous magnetic
field \cite{A2012PRA}, though this point is not important in the
present context.} along the paths shown in Fig. \ref{fig3path},
where the separations and recombinations of the path are obtained
through the so called \textquotedblleft
humpty-dumpty\textquotedblright\ problem \cite{schwinger-humpty,robert-humpty}%
. Weak interactions with quantum pointers can take place in the regions
$D,E,F,E^{\prime},F^{\prime},O$ and $O^{\prime}$ as indicated in the figure. A
final projective measurement takes place at time $t_{f}$ upon exiting the
interferometer by employing the beamsplitter SG2 in order to measure the spin
component along some direction $\mathbf{\hat{\phi}}$. The final post-selected
state is chosen to be
\begin{equation}
\left\vert \chi_{f}\right\rangle =\left\vert m_{f}\right\rangle \left\vert
\xi(t_{f})\right\rangle \equiv\sum_{k=-1}^{1}\left\langle m_{\alpha
}=k\right\vert \left.  m_{\phi}=+1\right\rangle \left\vert m_{\alpha
}=k\right\rangle \left\vert \xi(t_{f})\right\rangle \label{30}%
\end{equation}
with $\left\vert m_{f}\right\rangle \equiv\left\vert m_{\phi}=+1\right\rangle
$. The direction $\mathbf{\hat{\phi}}$ is chosen such that the following
condition holds:%
\begin{equation}
\sum_{k=-1,0}d_{k}(\alpha)\left\langle m_{f}\right\vert \left.  m_{\alpha
}=k\right\rangle =0.\label{condition1-1}%
\end{equation}

We can now compute the spatial projector weak values employing Eq.
(\ref{wvt}). Let $\Pi_{X}$ denote the spatial projector in the region $X,$
$\Pi_{X}=\left\vert \Gamma_{X}\right\rangle \left\langle \Gamma_{X}\right\vert
$ that can be taken to be a Gaussian encompassing at most the spatial extent
of the wavepacket, given by
\begin{equation}
\Gamma_{X}(\mathbf{r})=(\frac{2}{\pi\Delta^{2}})^{1/2}e^{-\left(
\mathbf{r}-\mathbf{r}_{X}\right)  ^{2}/\Delta^{2}}. \label{gam}%
\end{equation}
The results are (see the time labels on Fig. \ref{fig3path})
\begin{align}
t  &  =t_{1}\text{ : }\Pi_{E}^{w}=1\qquad\Pi_{F}^{w}=-1\label{ex1}\\
t  &  =t_{2}\text{ : }\Pi_{D}^{w}=1\qquad\Pi_{O}^{w}=0\label{ex2}\\
t  &  =t_{3}\text{ : }\Pi_{E^{\prime}}^{w}=1\qquad\Pi_{F^{\prime}}%
^{w}=-1\label{ex3}\\
t  &  =t_{4}\text{ : }\Pi_{O^{\prime}}^{w}=0, \label{ex4}%
\end{align}
assuming the projector width $\Gamma_{X}(\mathbf{r})$ [Eq. (\ref{gam})]
overlaps with the spatial wavefunction (otherwise the "ones" will be somewhat
smaller than 1, though the null weak values remain $0$). The computation of
these weak values is detailed in the Appendix.

Null weak values in Eqs. (\ref{ex1})-(\ref{ex4}) are obtained at $O$
and $O^{\prime}$. $\Pi_{O^{\prime}}^{w}=0$ can be understood from
the fact that the state vector going through $O^{\prime}$ is
orthogonal to the postselection state. The transition amplitude
$\left\langle \chi_{f}(t_{4})\right\vert \Pi_{O^{\prime}}\left\vert
\psi(t_{4})\right\rangle $ vanishes implying that the final state
$\left\vert \chi_{f}\right\rangle $ can therefore only be reached
via the upper path with $k=+1$ (going through $D$). Now the state
vector $\left\vert \psi(t_{4})\right\rangle $ going through
$O^{\prime}$ results from the superposition of the wavepackets
earlier localized at $E^{\prime}$ and $F^{\prime}$. Standard quantum
mechanics tells us that the overall transition amplitude
$\left\langle \chi_{f}(t_{3})\right\vert \left[
\Pi_{E^{\prime}}+\Pi_{F^{\prime}}\right]  \left\vert
\psi(t_{3})\right\rangle
$ vanishes but not the individual components $\left\langle \chi_{f}%
(t_{3})\right\vert \Pi_{E^{\prime}}\left\vert
\psi(t_{3})\right\rangle $ and $\left\langle
\chi_{f}(t_{3})\right\vert \Pi_{F^{\prime}}\left\vert \psi
(t_{3})\right\rangle $ and hence the weak values (\ref{ex3}) and
(\ref{ex4}) are non-null. The same reasoning applies to the weak
values $\Pi_{E}^{w}$ and $\Pi_{F}^{w}$ that do not vanish -- the
pointers placed at points $E$ and $F$ will therefore move -- while
$\Pi_{O}^{w}=0$ and the quantum pointer coupled to the system there
will not move. We interpret these results in Sec.\ \ref{interp}
below, but it should be noted that if the weak values $\Pi_{X}^{w}$
are taken to account for the particle being not there or there
according to whether the weak value is null or not, then we see that
our weakly coupled pointers detect a particle inside the inner loop
at $E^{\prime }$ and $F^{\prime}$ although no particle entered this
inner loop (as it wasn't detected by the pointer at $O$) and no
particle went out (as no particle was detected by the pointer at
$O^{\prime}$).

\subsubsection{Nested Mach-Zehnder\label{secmzi}}

\begin{figure}[tb]
\includegraphics[scale=0.15]{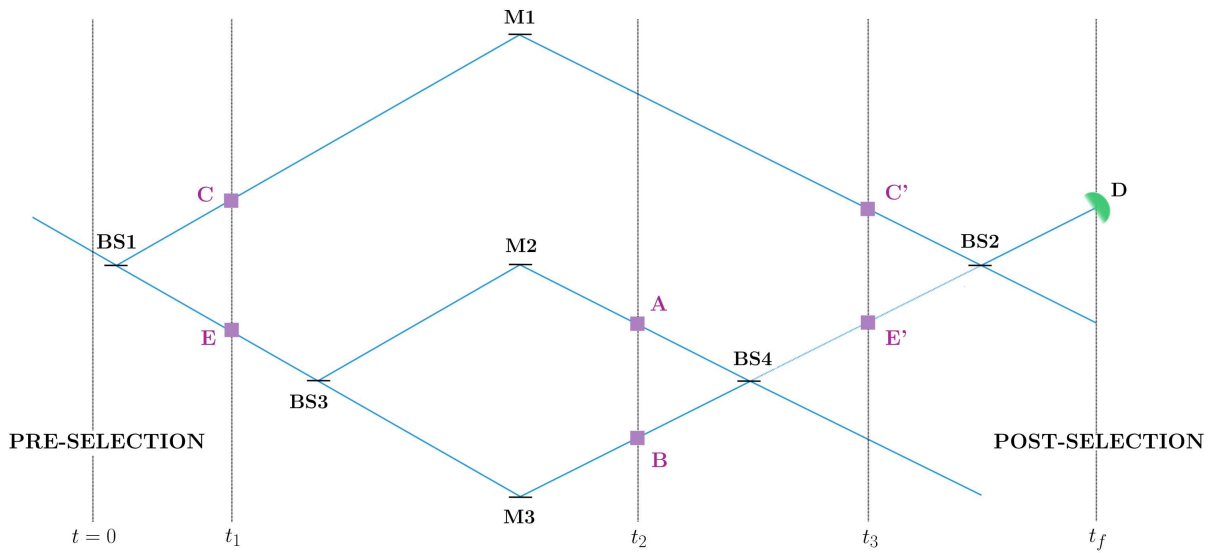}\caption{The nested Mach-Zehnder (MZ)
interferometer setup. For an appropriately postselected state, null weak
values are obtained at $E$ and $E^{\prime}$ (on the lower arm of the larger
MZ) but not inside the nested MZ (the projector weak values on arms $A$ and
$B$ are non-zero). }%
\label{figMZ}%
\end{figure}

The nested MZI example, introduced by Vaidman \cite{vaidman2013}, has been
amply reproduced and discussed in several papers
\cite{danan2013,zubairy2013,saldanha2014,vaidman-past,bartkiewicz2015,ferenczi2015,jordan-dove,sokolovskiMZI2016,griffiths2016,bula2016}%
, so we will only recall the main features. A photon enters a Mach-Zehnder
interferometer (arms $C$ and $E$ in Fig. \ref{figMZ}).\ A second MZI defining
paths $A$ and $B$, is placed on arm $E$ (labeled $E^{\prime}$ behind the
nested MZI). Postselection is defined by successful detection in port $D$. The
weak values are%
\begin{align}
t  &  =t_{1}\text{ : }\Pi_{C}^{w}=1\qquad\Pi_{E}^{w}=0\label{mz1}\\
t  &  =t_{2}\text{ : }\Pi_{A}^{w}=1\qquad\Pi_{B}^{w}=-1\label{mz2}\\
t  &  =t_{3}\text{ : }\Pi_{C^{\prime}}^{w}=1\qquad\Pi_{E^{\prime}}^{w}=0.
\label{mz3}%
\end{align}
As in the previous example, the detector appears to be reached only by photons
having taken arm $C$,on the ground that at $t=t_{3}$ previous to
postselection, $\Pi_{E^{\prime}}^{w}=0$. However inside the nested MZI on the
same arm, the weak values $\Pi_{A}^{w}$ and $\Pi_{B}^{w}$ are non-null
(pointers detect the photon's presence), although no photon can be detected
coming in or coming out since the weak values at $E$ and $E^{\prime}$ vanish.

\vspace{0.5cm} 

\section{Discussion}

\subsection{General remarks}

The main issue arising from the examples depicted in Figs. \ref{fig3path}
and\ \ref{figMZ} introduced in the preceding Section concerns the inference
that can be made on the past of a particle's motion based on the weak values.
A solution to this issue will depend on a thorough understanding of the weak
values (and more specifically on null weak values), and on being clear on the
underlying interpretational assumptions that are sometimes implicitly made
concerning the content of the standard formalism of quantum mechanics. The
salient feature that calls for an explanation -- irrespective of any stance
regarding the status of weak values -- is the fact that asymptotically weakly
coupled pointers are triggered when placed inside the \textquotedblleft
loops\textquotedblright\ seen in Figs. \ref{fig3path} and\ \ref{figMZ}, but
they are left intact (ie, they do not detect anything) when placed ahead of or
after the loop. We will not discuss here the explanations
\cite{zubairy2013,saldanha2014,bartkiewicz2015,ferenczi2015,bula2016} given
for the specific classical optics experiment reported in
Ref.\ \cite{danan2013}, that do not touch the fundamental aspects we are
focusing on in this work \footnote{Ref. \cite{zubairy2013} actually predates
the experiment, but the main argument in the present context is that in a
practical optics experiment attempts to simultaneously measure the weak values
given in Eqs. (\ref{mz1})-(\ref{mz3}) will result in leaks that will render
$\Pi_{E}^{w}$ and $\Pi_{E^{\prime}}^{w}$ non vanishing (we are assuming
instead that the couplings are sufficiently weak so that correlations between
weak pointers, that appear at second order in the coupling interactions, can
be neglected).}. From a fundamental standpoint, different approaches can be
considered, ranging from denying weak values have any bearing on the particle
properties (as properties hinge on a system being in an eigenstate of the
relevant observable), to assuming null weak values are a manifestation of some
novel underlying physics (like a wave coming from the future postselected
state). We will mostly focus here on analyzing how null weak values can be
interpreted.

\subsection{Interpretation of null weak values \label{interp}}

\subsubsection{Null weak values for projection operators \label{wvpo}}

As explained in Sec. \ref{nullwv}, a null weak value of a system observable
$A$ is a statement about a vanishing transition amplitude that can be inferred
from a quantum pointer coupled to $A$.\ If we are looking at the transition
amplitude of $\Pi_{X}\equiv\left\vert X\right\rangle \left\langle X\right\vert
$, then
\begin{equation}
\left\langle \chi(t_{w})\right\vert \Pi_{X}\left\vert \psi(t_{w})\right\rangle
=\left\langle \chi(t_{f})\right\vert U(t_{f},t_{w})\Pi_{X}U(t_{w}%
,t_{i})\left\vert \psi(t_{i})\right\rangle =0 \label{wvta}%
\end{equation}
is known from standard quantum mechanics to mean that the final
state $\left\vert \chi(t_{f})\right\rangle $ cannot be reached from
$\left\vert \psi(t_{i})\right\rangle $ by going through $X$.\ It is
important to stress that this is a statement concerning the
observable $\Pi_{X}$ (representing a physical property) and not the
wavefunction.\ An analogy with classical optics (as proposed in Ref.
\cite{saldanha2014} to describe the nested MZI of Sec. \ref{secmzi})
can at best be only partially useful, because although the classical
and quantum waves take all the paths inside the interferometers, the
classical electromagnetic wave is defined in physical space, whereas
the quantum wavefunction is defined over an abstract configuration
space and there is no consensus on its physical meaning
\footnote{The standard view is that the wavefunction doesn't refer
to a physical reality but is only a computational tool
\cite{A2002EJP}.}. This is the reason measurements in quantum
mechanics have a special status.

According to Eq. (\ref{wvta}), the postselected state cannot be
reached by the fraction of the system state coupled to the quantum
pointer
because that fraction evolved up to $t_f$, that is $U(t_{f},t_{w})\Pi_{X}U(t_{w}%
,t_{i})\left\vert \psi(t_{i})\right\rangle $, is orthogonal to the
postselected state. This property is not specific to weak
measurements. Indeed Eq. (\ref{finalpspos}) holds if the coupling is
strong\footnote{A strong coupling here does not imply a projective
measurement -- we are simply assuming the same unitary evolution as
per Eq. (\ref{corrs1}), but with a coupling strong enough to yield
orthogonal pointer states for each system eigenvalue.\ The
difference with an asymptotically weak coupling is that the
coherence properties of the system are spoiled by the orthogonality
of the entangled pointer states.}.\ Let us apply Eq.
(\ref{finalpspos}) with a strong coupling to the 3 path
interferometer for a quantum pointer placed at $O,$ initially in
state $\varphi_{O}(x,t_{i}).\ $The postselected state is given by
Eqs.\ (\ref{30})-(\ref{condition1-1}) and
$\Pi_{O}U(t_{2},t_{i})\left\vert \psi(t_{i})\right\rangle $ obtained
from Eq. (\ref{au}) is seen to be orthogonal to $\left\langle
\chi(t_{w})\right\vert .$ Therefore Eq. (\ref{finalpspos}) implies
that
\begin{equation}
\varphi_{O}(x,t_{f})=\varphi_{O}(x,t_{i}) \label{condo}%
\end{equation}
for each single run (for which postselection is obtained) -- the quantum
pointer has been left untouched by the strong coupling. This is unambiguously
taken to mean that the particle did not go through $O$.\ Applying the same
reasoning to a pointer strongly coupled to the particle at $D$ leads to
$\varphi_{D}(x,t_{f})\varpropto\varphi_{D}(x+g,t_{i}):$ for each run the
quantum pointer at $D$ acts as a detector that gets triggered, from which we
conclude that the particle took path $D$ (indeed, $\Pi_{D}U(t_{2}%
,t_{i})\left\vert \psi(t_{i})\right\rangle $ is \emph{not} orthogonal to
$\left\langle \chi(t_{2})\right\vert $). Now if the strongly coupled quantum
pointer is placed instead at $E^{\prime}$ or $F^{\prime}$ (or for that matter
at $E$ or $F$) there will be individual runs for which
\begin{equation}
\varphi_{E^{\prime}}(x,t_{f})\varpropto\varphi_{E^{\prime}}(x+g,t_{i})
\label{conde}%
\end{equation}
indicating that the particle was along path $E^{\prime}$. Having Eqs.
(\ref{condo}) and (\ref{conde}) is not seen as a contradiction because they
can never be realized jointly for strong interactions (in a Bohrian like
fashion, we would say that the conditions of the experiment are changed by
inserting strongly coupled pointers at different positions, so as a whole we
are not talking about the same physical situation).

In the asymptotically weak coupling limit however, all these conditions can be
realized jointly, because the weak interactions do not break the system
coherence. Arguably this cannot change the meaning of the transition
amplitudes: if $\left\langle \chi(t_{2})\right\vert \Pi_{O}\left\vert
\psi(t_{2})\right\rangle =0$ for a strong coupling implies that the system
having evolved from the initial state $\left\vert \psi(t_{i})\right\rangle $
cannot be found at $O$ when detected in state $\left\vert \chi(t_{f}%
)\right\rangle $, the same should hold for a weak coupling. The crucial
difference between strong and weak couplings concerns the system's state, not
the transition amplitude: strong interactions drives the system to an
eigenstate of the spatial projector, breaking the system coherence.\ The
eigenstate-eigenvalue link can then hold.\ This is not the case for weak
couplings, and this is precisely the reason the system coherence is not
modified and that weak values $\Pi_{O}^{w}=0$ and $\Pi_{E^{\prime}}^{w}=1$ can
be observed jointly. The bottom line is that the interpretation of a null weak
value as reflecting the absence of a system property in the region in which
the weak coupling took place hinges on one's stance concerning quantum
properties and the eigenvalue-eigenstate link (see Sec. \ref{sci}).

In our view, the fact that weakly coupled quantum pointers can detect whether
weak values are null or not are an indication that weak values can be regarded
as physical but they convey a different property ascription than the one
arising from the eigenstate-eigenvalue link. In the path integral approach, a
functional represents the value of a system property along each path
connecting the initial and final points, and the transition amplitude is
obtained by summing the functional over all the available interfering paths
(see Ch.\ 7\ of \cite{feynman-hibbs}). The null weak value at $O\ $in Fig.
\ref{fig3path} can be understood in this way -- the functional that takes
opposite values on paths $E$ and $F$ so that $\Pi_{E}^{w}=-\Pi_{F}^{w}$ is
summed at $O$ to yield a vanishing transition amplitude. From a quantum
perspective, there is nothing paradoxical in measuring a null weak value at
$O$ but not at $E$ and $E^{\prime}$: this appears as a consequence of taking
the superposition principle seriously. If the system cannot go through $O$ and
be detected in the postselected state, then we can say that \textquotedblleft
the particle was not there\textquotedblright\ provided \textquotedblleft
was\textquotedblright\ is employed in a liberal sense, because
\textquotedblleft the system is\textquotedblright\ is generally taken to mean
\textquotedblleft the system state is\textquotedblright, whereas here we are
discerning a particular particle property correlated with a transition to a
postselected state.

\subsubsection{Null weak values for general observables}

While our focus up to now was on null weak values for projectors, most of what
was written above holds also for null weak values of some general observable
$A,$ Eq. (\ref{wvta}) being replaced by%
\begin{equation}
\left\langle \chi(t_{w})\right\vert A_{X}\left\vert \psi(t_{w})\right\rangle
=\left\langle \chi(t_{f})\right\vert U(t_{f},t_{w})A_{X}U(t_{w},t_{i}%
)\left\vert \psi(t_{i})\right\rangle =0,
\end{equation}
the subscript $X$ indicating that $A$ is coupled to a quantum
pointer in region $X$ (ideally, we could write
$A_{X}\equiv\Pi_{X}A\Pi_{X}$ for a point like interaction at $X$).
The main difference is that projectors have a null eigenvalue,
rendering the connection between null weak values and eigenvalues of
projectors more straightforward than for observables that do not
possess a null eigenvalue. In particular the analogy made in Sec.
\ref{wvpo} above between strong and weak couplings does not work, as
a strong interaction couples the system eigenstates (with no null
eigenvalue) to orthogonal pointer states. But the interpretation
remains the same: the weak coupling  changes the tiny fraction of
the system state that couples to the quantum pointer into a state
that will evolve to be orthogonal to the postselected state. In case
of successful postselection, quantum correlations imply that the
property represented by $A$ will not couple to a pointer located at
$X$, and in this restricted sense, this property is ``not there''.
This conclusion is in line with Eq. (\ref{expvwv}) that tells us
that the expectation value of $A$ at time $t_{w}$ can be obtained at
time $t_{f}$ by measuring the
observable $B$, but disregarding the eigenstates $\left\vert b_{f}%
\right\rangle $ for which the transition amplitude $\left\langle
b_{f}\right\vert U(t_{f},t_{w})A\left\vert \psi(t_{w})\right\rangle $ vanishes.

To sum up, a null weak value should thus be understood as a statement
concerning the absence of the property represented by the observable in the
region in which the weak interaction took place, given the initial preparation
and conditioned on final postselection. It is important to emphasize that a
vanishing transition amplitude is to be associated with the absence of that
specific property of the system that coupled to the weak pointer. This is
sometimes forgotten when employing the \textquotedblleft weak
trace\textquotedblright\ criterion, as we now discuss.

\subsection{Inferring a particle's past\label{infering}}

\subsubsection{Weak trace criterion}

The "weak trace criterion" was defined in Ref. \cite{vaidman2013} as
indicating the whereabouts of a detected particle (in a fixed postselected
state) by looking at the weak trace left by the particle when locally coupled
to a quantum pointer. The coupling should be minimally disturbing, ie
sufficiently weak so that the coherence properties of the system are left
unaffected.\ Standard quantum mechanics tells us, as reviewed in Sec.
\ref{wv-f}, that the corresponding trace left on the quantum pointer's state
will precisely be the weak value of the system observable that coupled to the
pointer. Of course, a quantum particle is not a classical point-like object,
so we can expect to find simultaneous traces on different paths (like on the
two arms of an usual Mach-Zehnder interferometer). But according to the weak
trace criterion the particle was \emph{not} in regions where the projector
weak values vanished (and the relevant quantum pointers left intact). Now if
this criterion is endorsed, the illustrations given above in which a particle
leaves weak traces inside some inner loop, while no weak trace is left before
or after the loop, calls for an explanation.

Vaidman suggests this \textquotedblleft surprising\textquotedblright\ effect
can be explained naturally by adopting an interpretative framework combining
the two-state vector formalism (in which the weak values appear as the
effective interaction due to the overlap of a preselected state evolving
forward in time and a postselected state evolving backward in time) in the
context of the many-worlds model \cite{vaidman2013}. Alonso and Jordan
remarked \cite{jordan-dove} that adding prisms on the arms of the nested
interferometer in Fig. \ref{figMZ} did not change the weak values
(\ref{mz1})-(\ref{mz3}) but lead to detectable deflections at $E$ and
$E^{\prime}\ $. They wondered whether in a Wheeler-like fashion this effect
could not be interpreted as the photon leaving retroactively a trace at $E$
depending on the presence of a prism inserted after the photon has left arm
$E$ and entered the nested MZI.

Simpler explanations are available. First we remark that it is
perfectly possible (and it is generally the case) to have at some
point $X$ a vanishing spatial projector weak value in region
$\Pi_{X}^{w}$ while the weak value of another observable $A$ (like a
given spin component) measured at the same location is non-vanishing
($A^{w}\neq0$). This is straightforward to implement in the 3 path
interferometer by coupling at $O$ or $O^{\prime}$ an angular
momentum component $J_{\gamma}$ (where $\mathbf{\hat{\gamma}}$ can
be almost any arbitrary axis) to the quantum pointer;
$\Pi_{O}^{w}=0$ and $\Pi _{O^{\prime}}^{w}=0$ will still hold,
though the angular momentum weak values there $\left(
J_{\gamma}\right)  _{O}^{w}$ and $\left(  J_{\gamma}\right)  _{O^{\prime}}%
^{w}$ will be nonzero. In the nested MZI setup modified with prisms
\cite{jordan-dove}, it would arguably be simpler to write the
relevant photon observable related to the selective deflection
induced by the prism, and find that the corresponding weak values do
not vanish at $E^{\prime}$ and $E^{\prime}$. Hence the weak trace
criterion should be therefore be employed with reference to a
specific system property. If we specify that we are inferring the
particle's past trajectory, since a trajectory is defined by the
space-time points $\{t_{k},\mathbf{r}(t_{k})\}$, the relevant weak
measurements are those related to the sole system position, and
involve indeed the projection operators.

This brings us to the second point: a quantum particle is not a
classical object (hence not even a particle in this sense).
Inferring a particle's past (and not only its past trajectory)
should then involve the different properties that can be measured.
Weak values of different observables will vanish at different
locations. While detecting different properties in alternative
locations would be startling for a classical particle, this is not
so for an evolving quantum system envisaged as an extended
undulatory entity whose local properties depend on interfering
paths.

\subsubsection{Strong trace criterion\label{sci}}

We term here \textquotedblleft strong trace criterion\textquotedblright\ the
scheme according to which a quantum particle's past only makes sense when
based on the eigenstate-eigenvalue link. This is remarkably the case of the
Consistent Histories approach, whose starting point is to define a property
from eigenvectors spanning the corresponding Hilbert space
subspaces.\ Griffiths has recently given a Consistent Histories (CH) account
of the nested MZI problem \cite{griffiths2016}. CH asserts that attempting to
give an account of the particle's presence inside the inner MZI is
meaningless: the history family in which arms $A$ and $B$ of the inner MZI
would be treated as mutually exclusive is inconsistent. This is to be expected
whenever properties are grounded on assigning probabilities, and the CH
framework precisely pinpoints what type of histories can describe an evolving
quantum system and why two histories may be incompatible on this ground. While
there is no place for weak measurements in the CH\ approach (given that weak
measurements do not abide by the eigenstate-eigenvalue link), it would be
instructive to see how CH explains the existence of weakly coupled pointers
that measure quantities proportional to transition amplitudes. Unfortunately
this is not done in Ref.\ \cite{griffiths2016}, where instead of weak
measurements as introduced in Sec. \ref{wv-f}, strong interactions with a weak
probability are discussed (the implications are examined in
\cite{vaidman-com2016}).

Employing a totally different framework also based ultimately on obtaining
probabilities as specified by the eigenstate-eigenvalue link, Sokolovski
\cite{sokolovskiMZI2016} does attempt to give a meaning to the weakly coupled
pointers. In his view a path is real if a probability for taking a path can be
obtained, but a path is virtual if only a transition amplitude can be attached
to it. A strongly coupled meter creates real paths, while in the limit of
small interactions a weakly coupled pointer picks up a \textquotedblleft
relative path amplitude\textquotedblright\ that has no bearing on the real
interactions that have taken place.\ A vanishing transition amplitude at $X$
is then only relevant insofar as it indicates that a single standard strong
pointer inserted at $X$ would not detect the particle there, but according to
\cite{sokolovskiMZI2016} it is meaningless to make any assertion concerning
the property of the system if interferences are not lifted by a strong
coupling that will end up projecting the pointer to a state associated with a
given system eigenstate.

The \textquotedblleft strong trace criterion\textquotedblright\ fits well with
the conventional view in which a property (represented by an observable) can
only be ascribed to a quantum system when it is in an eigenstate of that
observable.\ But from the start, the \textquotedblleft strong trace
criterion\textquotedblright\ discards any possibility to infer a property from
protocols implementing non-destructive weak interactions. By restricting
quantum properties ascription to changes of the state vector, the
\textquotedblleft strong trace criterion\textquotedblright\ has difficulties
in giving a significance to the output of weakly coupled pointers that do not
change the state of the system but give an indication of the value of an
observable correlated with a detection in a postselected state. Indeed, such
pointers, that can be experimentally observed, are then given a counterfactual
significance (if a projective measurement would have been made instead then
the result indicated by that particular weak pointer would have been
obtained), a rather peculiar stance.

\section{Conclusion}

In this work we have analyzed the properties and meaning of null weak values
in the context of inferring the past of a quantum particle from interactions
of the system with weakly coupled pointers. A null weak value of an observable
$A$ obtained at some location $X$ means that the system property represented
by $A$ cannot be found at $X$ and detected in the postselected state. The past
of a quantum particle can be inferred by taking into account all of its
observables, not only spatial projectors. The fact that discontinous traces of
a given property can be experimentally observed from weakly coupled pointers
seems to be an indication that\thinspace the wavefunction superposition is
related to a physical phenomenon, rather than being a mere computational artifact.

\appendix

\section{Weak values in the 3-path interferometer}

We detail here the computation of the weak values for the 3 path
interferometer described in Sec.\ \ref{3ex}. As an example, let us give the
calculation for the weak values at $t=t_{2}.$ We have by the very definition
Eq. (\ref{wvt})%
\begin{equation}
\Pi_{D}^{w}=\frac{\left\langle \chi_{f}(t_{f})\right\vert U(t_{f},t_{2}%
)\Pi_{D}U(t_{2},t_{i})\left\vert \psi_{i}\right\rangle }{\left\langle \chi
_{f}(t_{f})\right\vert U(t_{f},t_{2})U(t_{2},t_{i})\left\vert \psi
_{i}\right\rangle }.
\end{equation}
Then keeping in mind that $\Pi_{D}\left\vert \xi_{k}(t_{2})\right\rangle =0 $
for $k=0,-1$ Eq. (\ref{27}) leads to
\begin{equation}
\Pi_{D}^{w}=\frac{\left\langle \xi_{f}(t_{f})\right\vert U(t_{f},t_{2})\Pi
_{D}\left\vert \xi_{k=+1}(t_{2})\right\rangle d_{1}(\alpha)\left\langle
m_{f}\right\vert \left.  m_{\alpha}=1\right\rangle }{\sum_{k=-1}^{1}%
d_{k}(\alpha)\left\langle m_{f}\right\vert \left.  m_{\alpha}=k\right\rangle }%
\end{equation}
that simplifies given our choice of $\left\vert m_{f}\right\rangle ,$
encapsulated by the condition (\ref{condition1-1}) to
\begin{equation}
\Pi_{D}^{w}=\left\langle \xi(t_{f})\right\vert U(t_{f},t_{2})\Pi_{D}\left\vert
\xi_{A}(t_{2})\right\rangle \approx1.
\end{equation}
For the weak value in the region $O$ we have
\begin{equation}
\Pi_{O}^{w}=\frac{\left\langle \chi_{f}(t_{f})\right\vert U(t_{f},t_{2}%
)\Pi_{O}U(t_{2},t_{i})\left\vert \psi_{i}\right\rangle }{\left\langle \chi
_{f}(t_{f})\right\vert U(t_{f},t_{2})U(t_{2},t_{i})\left\vert \psi
_{i}\right\rangle }.
\end{equation}
Following Eq. (\ref{27}), $U(t_{2},t_{i})\left\vert \psi_{i}\right\rangle $ is
of the form
\begin{equation}
U(t_{2},t_{i})\left\vert \psi_{i}\right\rangle =d_{1}(\alpha)\left\vert
m_{\alpha}=+1\right\rangle \left\vert \xi_{D}(t_{2})\right\rangle
+\sum_{k=-1,0}d_{k}(\alpha)\left\vert m_{\alpha}=k\right\rangle \left\vert
\xi_{O}(t_{2})\right\rangle \label{au}%
\end{equation}
and $\Pi_{O}\left\vert \xi_{D}(t_{2})\right\rangle $ vanishes (since there is
no spatial overlap between $\left\vert \Gamma_{O}\right\rangle $ and
$\left\vert \xi_{D}(t_{2})\right\rangle $). The weak value becomes
\begin{equation}
\Pi_{O}^{w}=\frac{\left\langle \xi(t_{f})\right\vert U(t_{f},t_{2})\Pi
_{O}\left\vert \xi_{O}(t_{2})\right\rangle }{\left\langle \chi_{f}%
(t_{f})\right\vert U(t_{f},t_{2})U(t_{2},t_{i})\left\vert \psi_{i}%
\right\rangle }\left[  \sum_{k=-1,0}d_{k}(\alpha)\left\langle m_{f}\right\vert
\left.  m_{\alpha}=k\right\rangle \right]  =0;
\end{equation}
indeed, the square bracket in this equation vanishes, since this is precisely
the condition (\ref{condition1-1}) imposed for the postselection state.

The other weak values given in Eqs. (\ref{ex1})-(\ref{ex4}) are computed in
the same way.

\bibliographystyle{abbrvNoTitle}
\bibliography{bibliographyO}

\end{document}